\documentclass{PoS}

\usepackage{amsmath}
\usepackage{graphicx}

\def\rmO{{\rm O}}

\def\proof{\noindent{\sl Proof:}\kern0.6em}

\def\frac#1#2{\hbox{$#1\over#2$}}
\def\dual{\mathstrut^*\kern-0.1em}

\def\lvec#1{\setbox0=\hbox{$#1$}
    \setbox1=\hbox{$\scriptstyle\leftarrow$}
    #1\kern-\wd0\smash{
    \raise\ht0\hbox{$\raise1pt\hbox{$\scriptstyle\leftarrow$}$}}
    \kern-\wd1\kern\wd0}
\def\lrvec#1{\setbox0=\hbox{$#1$}
    \setbox1=\hbox{$\scriptstyle\leftarrow$}
    #1\kern-\wd0\smash{
    \raise\ht0\hbox{$\raise1pt\hbox{$\scriptstyle\leftrightarrow$}$}}
    \kern-\wd1\kern\wd0}
\def\rvec#1{\setbox0=\hbox{$#1$}
    \setbox1=\hbox{$\scriptstyle\rightarrow$}
    #1\kern-\wd0\smash{
    \raise\ht0\hbox{$\raise1pt\hbox{$\scriptstyle\rightarrow$}$}}
    \kern-\wd1\kern\wd0}

\def\Dslash{{D\kern-7pt\slash}}

\def\nabstar#1{\nabla\kern-0.5pt\smash{\raise 4.5pt\hbox{$\ast$}}
               \kern-4.5pt_{#1}}

\def\drvstar#1{\partial\kern-0.5pt\smash{\raise 4.5pt\hbox{$\ast$}}
               \kern-5.0pt_{#1}}

\def\momp#1#2{
    \setbox0=\hbox{${#1}$}\setbox1=\hbox{${#1}_{#2}$}
    {#1}_{#2}\kern-\wd1\kern\wd0
    \smash{\raise4.5pt\hbox{$\scriptscriptstyle +$}}}
\def\momm#1#2{
    \setbox0=\hbox{${#1}$}\setbox1=\hbox{${#1}_{#2}$}
    {#1}_{#2}\kern-\wd1\kern\wd0
    \smash{\raise4.5pt\hbox{$\scriptscriptstyle -$}}}
\def\mompm#1#2{
    \setbox0=\hbox{${#1}$}\setbox1=\hbox{${#1}_{#2}$}
    {#1}_{#2}\kern-\wd1\kern\wd0
    \smash{\raise4.5pt\hbox{$\scriptscriptstyle \pm$}}}
\def\smomp#1#2{
    \setbox0=\hbox{${#1}$}\setbox1=\hbox{${#1}_{#2}$}
    {#1}_{#2}\kern-\wd1\kern\wd0
    \smash{\raise3pt\hbox{$\scriptscriptstyle +$}}}
\def\smomm#1#2{
    \setbox0=\hbox{${#1}$}\setbox1=\hbox{${#1}_{#2}$}
    {#1}_{#2}\kern-\wd1\kern\wd0
    \smash{\raise3pt\hbox{$\scriptscriptstyle -$}}}
\def\smompm#1#2{
    \setbox0=\hbox{${#1}$}\setbox1=\hbox{${#1}_{#2}$}
    {#1}_{#2}\kern-\wd1\kern\wd0
    \smash{\raise3pt\hbox{$\scriptscriptstyle \pm$}}}

\def\rhoprime{\rho\kern1pt'}
\def\rhobar{\bar{\rho}}
\def\rhobarprime{\rhobar\kern1pt'}
\def\rhobartilde{\kern2pt\tilde{\kern-2pt\rhobar}}
\def\rhobartildeprime{\kern2pt\tilde{\kern-2pt\rhobar}\kern1pt'}

\def\zetabar{\overline{\zeta}}
\def\zetaprime{\zeta\kern1pt'}
\def\zetabarprime{\zetabar\kern1pt'}
\def\zetar{\zeta_{\raise-1pt\hbox{\rm R}}}
\def\zetabarr{\zetabar_{\raise-1pt\hbox{\rm R}}}

\def\phiimpr{\phi_{\kern0.5pt{\rm I}}}

\def\diracstar#1#2{
    \setbox0=\hbox{$\gamma$}\setbox1=\hbox{$\gamma_{#1}$}
    \gamma_{#1}\kern-\wd1\kern\wd0
    \smash{\raise4.5pt\hbox{$\scriptstyle#2$}}}

\def\csw{c_{\rm sw}}

\def\ct{c_{\rm t}}

\def\Nf{N_{\rm f}}

\def\opprime#1{\setbox0=\hbox{${\cal O}$}\setbox1=\hbox{${\cal O}_{\rm #1}$}
    {\cal O}_{\rm #1}\kern-\wd1\kern\wd0
    \smash{\raise4.5pt\hbox{\kern1pt$\scriptstyle\prime$}}\kern1pt}

\def\ophatprime#1{\setbox0=\hbox{$\widehat{\cal O}$}
    \setbox1=\hbox{$\widehat{\cal O}_{\rm #1}$}
    \widehat{\cal O}_{\rm #1}\kern-\wd1\kern\wd0
    \smash{\raise4.5pt\hbox{\kern1pt$\scriptstyle\prime$}}\kern1pt}

\def\bopprime#1{\setbox0=\hbox{${\cal O}$}\setbox1=\hbox{${\cal O}_{\rm #1}$}
    {\cal L}_{\rm #1}\kern-\wd1\kern\wd0
    \smash{\raise4.5pt\hbox{\kern1pt$\scriptstyle\prime$}}\kern1pt}

\def\blagprime#1{\setbox0=\hbox{${\cal B}$}\setbox1=\hbox{${\cal B}_{#1}$}
    {\cal B}_{#1}\kern-\wd1\kern\wd0
    \smash{\raise5.2pt\hbox{\kern1pt$\scriptstyle\prime$}}\kern1pt}

\def\mbar{\kern1pt\overline{\kern-1pt m\kern-1pt}\kern1pt}

\def\msbar{{\rm \overline{MS\kern-0.05em}\kern0.05em}}

\def\blackboardrrm{\mathchoice
{\rm I\kern-0.21 em{R}}{\rm I\kern-0.21 em{R}}
{\rm I\kern-0.19 em{R}}{\rm I\kern-0.19 em{R}}}
 
\def\blackboardzrm{\mathchoice
{\rm Z\kern-0.32 em{Z}}{\rm Z\kern-0.32 em{Z}}
{\rm Z\kern-0.28 em{Z}}{\rm Z\kern-0.28 em{Z}}}
 
\def\blackboardh{\mathchoice
{\rm I\kern-0.14 em{H}}{\rm I\kern-0.14 em{H}}
{\rm I\kern-0.11 em{H}}{\rm I\kern-0.11 em{H}}}
 
\def\blackboardp{\mathchoice
{\rm I\kern-0.14 em{P}}{\rm I\kern-0.14 em{P}}
{\rm I\kern-0.11 em{P}}{\rm I\kern-0.11 em{P}}}
 
\def\blackboardt{\mathchoice
{\rm T\kern-0.52 em{T}}{\rm T\kern-0.52 em{T}}
{\rm T\kern-0.40 em{T}}{\rm T\kern-0.40 em{T}}}

\def\deltaoneprime{\Delta\kern-1.0pt
    \smash{\raise 4.5pt\hbox{$\scriptstyle\prime$}}
    \kern-1.5pt_{1}}

\def\diag{{\rm diag}}

\def\delstar#1{\Delta\kern-1.0pt\smash{\raise 4.5pt\hbox{$\ast$}}
               \kern-4.0pt_{#1}}

\def\nabstar#1{\nabla\kern-0.5pt\smash{\raise 4.5pt\hbox{$\ast$}}
               \kern-4.5pt_{#1}}

\def\lnabstar#1{\overleftarrow{\nabla}\kern-0.5pt\smash
             {\raise 4.5pt\hbox{$\ast$}}\kern-4.5pt_{#1}}
\def\cdev#1{D\kern-0.2pt\smash{\raise 4.2pt
            \hbox{$\scriptstyle\phantom{\ast}$}}
            \kern-4.8pt_{#1}}
\def\cdevstar#1{D\kern-0.2pt\smash{\raise 4.2pt
                \hbox{$\scriptstyle\ast$}}
                \kern-4.8pt_{#1}}

\newcommand{\Dsl}{D \kern-.65em/}

\newcommand{\be}{\begin{equation}}
\newcommand{\ee}{\end{equation}}
\newcommand{\bi}{\begin{itemize}}
\newcommand{\ei}{\end{itemize}}
\newcommand{\bes}{\begin{eqnarray}}
\newcommand{\ees}{\end{eqnarray}}
\newcommand{\bea}{\begin{eqnarray}}
\newcommand{\eea}{\end{eqnarray}}
\newcommand{\bean}{\begin{eqnarray*}}
\newcommand{\eean}{\end{eqnarray*}}

\title{Perturbative lattice artefacts in the SF coupling for technicolor-inspired models}

\ShortTitle{Perturbative lattice artefacts in the SF coupling}

\author{\speaker{Stefan Sint} and Pol Vilaseca\\
        School of Mathematics, Trinity College, Dublin 2, Ireland\\
        E-mail: \email{sint@maths.tcd.ie, pol@maths.tcd.ie}}

\abstract{Viable candidate theories for electroweak dynamical symmetry breaking 
are expected to show (near) conformal behaviour in order to accommodate current 
phenomenological constraints. In principle, renormalisation group studies using 
finite volume renormalisation schemes are well-suited to verify 
this property in a given model. The most practical schemes are based on the 
Schr\"odinger functional (SF), but suffer from potentially large O(a) effects. 
Some care has to be taken to remove these effects and to set up a scheme 
where cutoff effects are small and under control. 
We here take a step in this direction by analysing various set-ups 
for the SF coupling at one-loop order in perturbation theory.
}

\FullConference{ The XXIX International Symposium on Lattice Field Theory - Lattice 2011\\
July 10-16, 2011\\
Squaw Valley, Lake Tahoe, California}

\begin{document}

\section{Introduction}

The prospect of discovering the underlying dynamics of electroweak symmetry 
breaking at the LHC has triggered new efforts to check the viability of technicolor-inspired
models of dynamical symmetry breaking~(see e.g.~\cite{Andersen:2011yj}). It is widely believed that
a viable model should show conformal or near conformal behaviour over some range of scales.
Such behaviour is expected for some QCD inspired models with certain values of $(\Nf,N)$, 
where $\Nf$ denotes the number of fermion flavours and $N$ the number of colours. Furthermore, one
needs to choose the representation of the gauge group for the fermions. We
will here focus on the example of $\Nf=2$ flavours in the two-index symmetric (sextet) representation of
the colour group SU(3).

Non-perturbative studies of the renormalization group evolution of the coupling
provide a direct way to test for (near) conformal behaviour.  Finite volume renormalization schemes 
based on the Schr\"odinger functional (SF)~\cite{Luscher:1992an,Sint:1993un} have become popular in 
this context~(cf.~\cite{DelDebbio:2010zz} for a recent review). 
However, some care has to be taken to distinguish universal continuum properties from lattice artefacts.
Before embarking on a non-perturbative study it is thus advisable to test any proposed 
framework in perturbation theory, where the continuum limit is known and 
the size of lattice artefacts can be assessed. 

In this contribution we propose a closer look at the set-up of the SF
with Wilson fermions. As is well-known from lattice QCD, it is important to remove
lattice effects linear in $a$, which originate both from the time boundaries and from 
the bulk~\cite{Luscher:1996sc}. The bulk O($a$) effects can be completely eliminated by 
non-perturbative Symanzik improvement~\cite{Luscher:1996ug}, 
and for fermions in higher SU($N$) representations  this has been 
attempted in ref.~\cite{Karavirta:2011mv}. An alternative consists in using the chirally 
rotated SF ($\chi$SF) \cite{Sint:2010eh,Sint:2010xy}, 
which implements the mechanism of automatic O($a$) improvement. We here compare both options, 
and our study also provides a further test of the $\chi$SF in perturbation theory.

Even if the leading O($a$) effects are eliminated, the size of the remaining higher lattice
artefacts in the coupling can be large, and in some cases the asymptotic O($a^2$) behaviour
sets in rather late, beyond the typical lattice sizes accessible to simulations~\cite{Sommer:1997jg}.
While we only provide details for fundamental and the sextet representations of SU(3),
our observations are generic and thus also relevant for gauge group SU(2) 
and fermions in other other non-fundamental representations~\cite{Karavirta,progress}.

This writeup is organized as follows: after a short reminder of the basic definitions 
of the SF coupling we discuss the parameterization of cutoff effects in the step-scaling function
and criteria for assessing their size. We then discuss the perturbative one-loop data
before we draw our conclusion and outline future directions.

\section{The SF coupling}

The Schr\"odinger functional provides us with a mass-independent finite volume scheme for the 
coupling. It is defined through the variation of a colour-electric background field $B_\mu$
with respect to a parameter $\eta$. Denoting its effective action by 
$\Gamma[B]$, the coupling is defined by~\cite{Luscher:1993gh,Sint:1995ch},
\begin{equation}
   \bar{g}^2(L) = \left.\dfrac{\partial_\eta \Gamma_0[B]}{\partial_\eta \Gamma[B]}\right\vert_{\eta=0}, \qquad
   \Gamma[B]\; \stackrel{g_0\rightarrow 0}{\sim}\; \frac{1}{g_0^2}\Gamma_0[B] + \Gamma_1[B] +\ldots.
\end{equation}
At one-loop order in perturbation theory its relation to the bare coupling $g_0$ is of the form,
\begin{equation}
   \bar{g}^2(L) = g_0^2 + p_1(L/a) g_0^4 + \ldots, \qquad p_1(L/a) = p_{1,0}(L/a) + \Nf p_{1,1}(L/a).
\end{equation}
Although the SF coupling is an observable in numerical simulations, it is
expensive to measure and simulations with Wilson fermions are typically limited to lattices 
with $L/a \le 16$. It is therefore important to find SF schemes where cutoff effects are as small as 
possible and where the asymptotic regime is reached already on such relatively small lattices.

\section{Parameterization of cutoff effects}

In the traditional analysis of lattice QCD data the quantity of interest is the step-scaling 
function (SSF),
\begin{equation}
   \Sigma(u,a/L) = \bar{g}^2(2L)\vert_{u=\bar{g}^2(L)} = u + \Sigma_1(a/L) u^2 + \rmO(u^3).
\end{equation}
It has a universal continuum limit, 
\begin{equation}
   \lim_{a\rightarrow 0} \Sigma(u,a/L) = \sigma(u) = u + \sigma_1 u^2 + \rmO(u^3),\qquad \sigma_1= 2b_0 \ln(2),
\end{equation}
where the leading term contains the one-loop coefficient of the $\beta$-function,
\begin{equation}
   b_0 = b_{0,0}+ \Nf b_{0,1} = \dfrac{1}{16\pi^2}\left(\dfrac{11N}{3}-\Nf\dfrac{4}{3} T_R\right),
\end{equation}
with $T_R = 1/2, N, (N+2)/2$ for $R$ denoting the fundamental, adjoint and two-index symmetric representations,
respectively. We now define relative deviations from  pure gauge and fermionic continuum coefficients in 
$\sigma_1= \sigma_{1,0} + \Nf \sigma_{1,1}$:
\begin{equation}
   \delta_{1,0}(a/L) = \dfrac{\Sigma_{1,0}(a/L)-\sigma_{1,0}}{\sigma_{1,0}},\qquad
   \delta_{1,1}(a/L) = \dfrac{\Sigma_{1,1}(a/L)-\sigma_{1,1}}{\sigma_{1,1}}.
\end{equation}
How large can we allow the cutoff effects to be? It is instructive to study this question 
at fixed lattice size: a relative deviation  of $\delta_{1,1}= 100 \%$ 
could then be mistaken for getting $\Nf$ wrong by a factor 2, since
\begin{equation}
  \Nf\Sigma_{1,1} = (1+ \delta_{1,1}) \Nf \sigma_{1,1} \approx 2\Nf\sigma_{1,1}.
\end{equation}
Hence, with all parameters fixed, requiring $|\delta_{1,1}(a/L)| \ll 1$ seems reasonable and sets a lower limit on $L/a$.
For comparison, the relative deviation in the pure SU(3) gauge theory on the smallest lattices is given by
$\delta_{1,0}(1/4)= -20\%$, $\delta_{1,0}(1/6)= -8\%$  and $\delta_{1,0}(1/8)= -4\%$, 
and thus quickly approaches zero. It seems natural to require a similar behaviour for the contribution from the fermions.

\section{O($a$) vs. O($a^2$) asymptotic behaviour}

O($a$) effects in the SSF with the standard SF originate from both the bulk and the boundaries.
Boundary O($a$) effects can be cancelled by the local boundary counterterms $\propto \ct,\tilde{\ct}$~\cite{Luscher:1996sc}.
The only bulk counterterm is the clover term with coefficient $\csw$. With the $\chi$SF 
only boundary O($a$) counterterms are required ($\propto \ct, d_s$)~\cite{Sint:2010eh}.
The price to be paid for this advantage consists in the tuning of $z_f$, 
a dimension-3 boundary counterterm, in addition to the usual tuning of the bare mass 
to its critical value $m_{\rm cr}$. In our perturbative context the critical fermion mass  
and $z_f$ are set to their tree-level values, $m_{\rm cr}=0$ and $z_f=1$.
We also assume that tree-level boundary O($a$) improvement is implemented, 
i.e.~we set the tree-level coefficients to $\ct^{(0)}=1$, $\tilde{\ct}^{(0)}=1$, $d_s^{(0)}=0.5$.

\begin{figure}
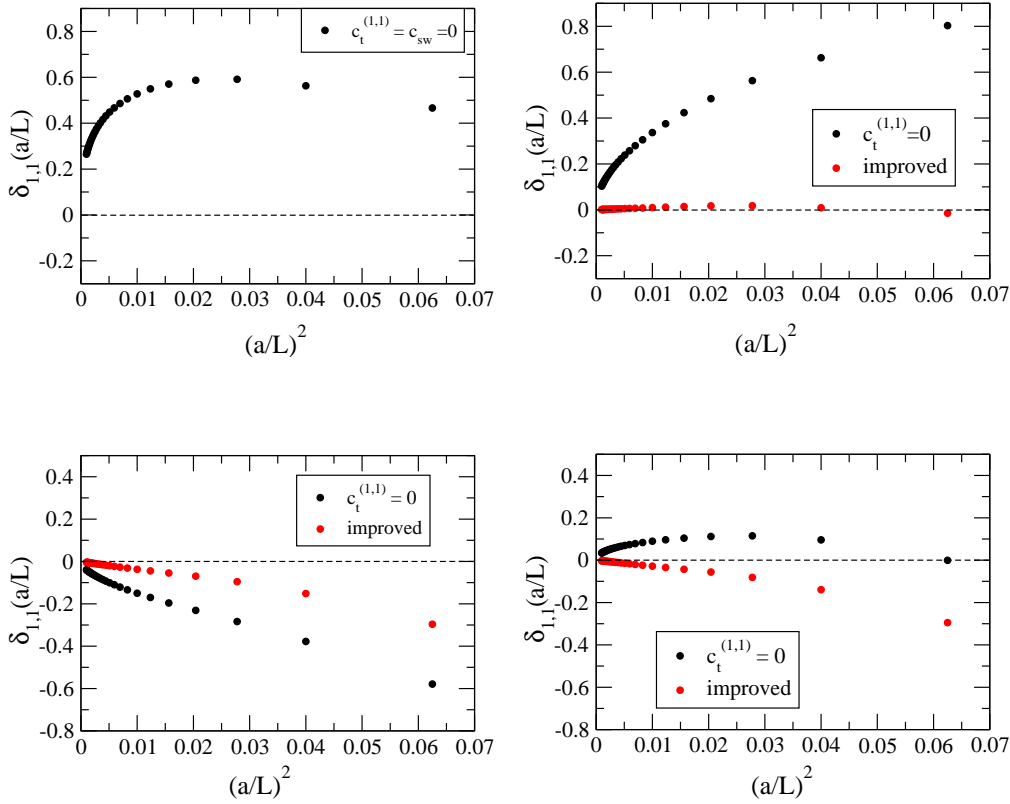

\vskip 2.5ex
\includegraphics[width=.43\textwidth]{./plots/delta11_SF_SU3f_csw0}\hskip 2ex
\includegraphics[width=.43\textwidth]{./plots/delta11_SF_SU3f_csw1.eps}
\vskip 7ex
\includegraphics[width=.43\textwidth]{./plots/delta11_XSF_SU3f_csw0.eps}\hskip 2ex
\includegraphics[width=.43\textwidth]{./plots/delta11_XSF_SU3f_csw1.eps}
\caption{Lattice artefacts $\delta_{1,1}(a/L)$ for fermions in the fundamental representation of SU(3), regularised with 
the standard SF (upper panels) and the $\chi$SF (lower panels), respectively.
The 2 panels on the left (right) show data with $\csw=0$ ($\csw=1$).}
\label{fig1}
\end{figure}

\section{Results}

The asymptotic behaviour of the fermionic one-loop coefficient can be parameterized as follows:
\begin{equation}
   p_{1,1}(L/a)  \stackrel{a/L \rightarrow 0}{\sim} r_0 + s_0\ln(L/a) + \dfrac{a}{L} \Bigl(r_1 + s_1 \ln(L/a)\Bigr) + \rmO(a^2).
\end{equation}
Independently of the regularisation one expects that $s_0$ is given by $s_0 = 2b_{0,1}$. 
The coefficient $r_1$ is cancelled by the correct choice of $\ct^{(1,1)}$ and $s_1$ is either 
proportional to $\csw-1$ (standard SF) or expected to vanish ($\chi$SF).
We now consider the following cases: 
\begin{enumerate}
\item standard SF, unimproved ($\ct^{(1,1)}=0$ and $\csw=0$); 
\item standard SF, improved in the bulk ($\csw=1$), but not at boundaries ($\ct^{(1,1)}=0$);
\item standard SF, fully improved;
\item $\chi$SF, unimproved at the boundaries ($\ct^{(1,1)}=0$) with either $\csw=0$ or $\csw=1$;
\item $\chi$SF, fully improved, with either $\csw=0$ or $\csw=1$.
\end{enumerate}
With fermions in the fundamental representations we see that the cutoff effects are
essentially zero in the standard SF, once improvement both in the bulk and at the boundaries is
implemented (figure~\ref{fig1}). This smallness is likely to be a kinematical accident for this choice of background
field and parameters. What is striking, however, is the large size of the effects cancelled by both $\csw$ and
$\ct^{(1,1)}$. In the $\chi$SF the situation is as expected: the cutoff effects behave asymptotically as O($a^2$)
once the boundary improvement is implemented. We note that the boundary cutoff effect is smaller here
and the asymptotic O($a^2$) behaviour is indeed observed (figure~\ref{fig1}).

\begin{figure}
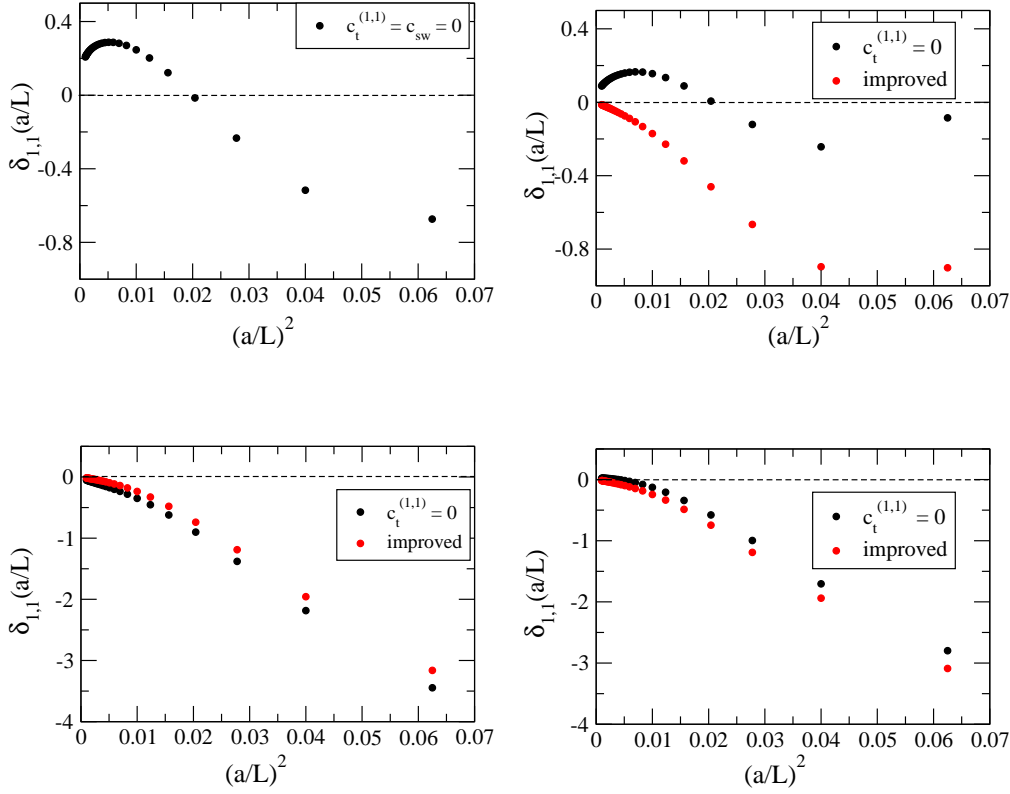

\vskip 2.5ex
\includegraphics[width=.43\textwidth]{./plots/delta11_SF_SU3s_csw0.eps}\hskip 2ex
\includegraphics[width=.43\textwidth]{./plots/delta11_SF_SU3s_csw1.eps}
\vskip 7ex
\includegraphics[width=.43\textwidth]{./plots/delta_11_SU3_symm_ds05_csw0.eps}\hskip 2ex
\includegraphics[width=.43\textwidth]{./plots/delta_11_SU3_symm_ds05_csw1.eps}
\caption{The same as figure~1, for the 2-index symmetric (sextet) representation of SU(3).}
\label{fig2}
\end{figure}

The situation changes drastically if the fundamental representation is
replaced by the sextet representation. In this case
the cutoff effects are enormous, as illustrated in figure~\ref{fig2}.
The situation is bad enough with the standard SF and quite a bit worse with the $\chi$SF.
This is clearly a disaster, and it is obvious that neither framework 
could be used to extract any sensible continuum results based on data for only
a few relatively small lattices.

\subsection{A possible cure: weaken the background field?}

\begin{figure}
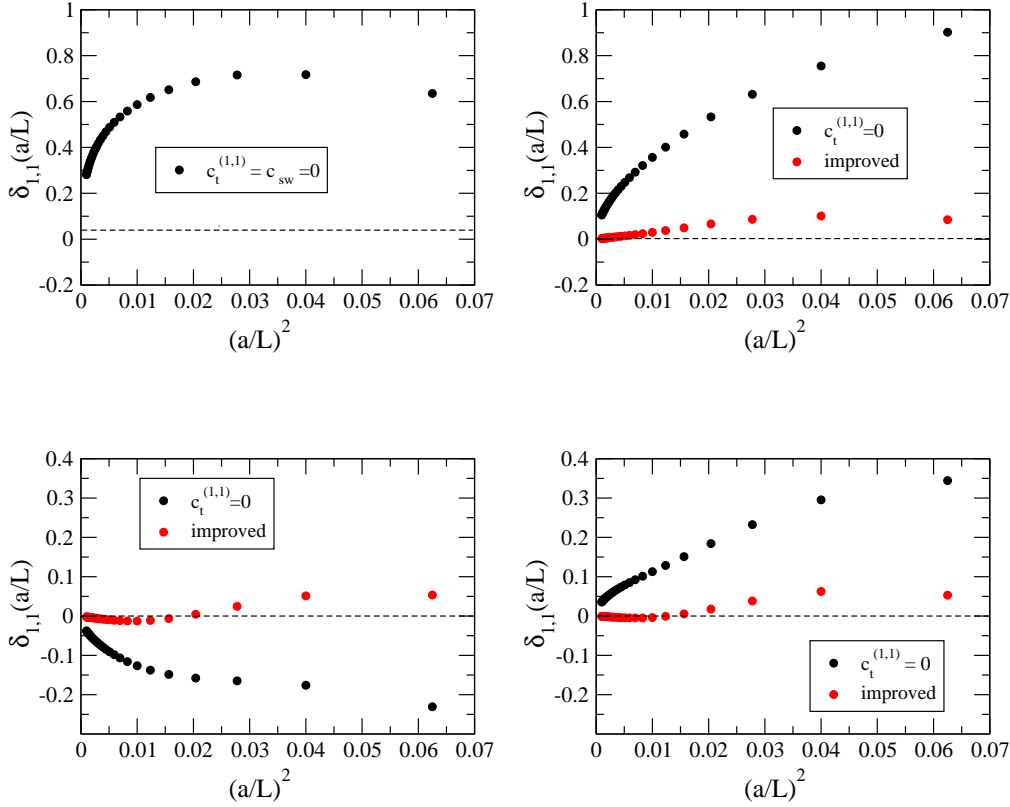

\vskip 2.5ex
\includegraphics[width=.43\textwidth]{./plots/delta11_SF_SU3s_csw0_BF2.eps}\hskip 2ex
\includegraphics[width=.43\textwidth]{./plots/delta11_SF_SU3s_csw1_BF2.eps}
\vskip 7ex
\includegraphics[width=.43\textwidth]{./plots/delta11_XSF_SU3s_csw0_BF2.eps}\hskip 2ex
\includegraphics[width=.43\textwidth]{./plots/delta11_XSF_SU3s_csw1_BF2.eps}
\caption{The same as figure~2 with a background field $B$ which is weaker by a factor $1/2$.}
\label{fig3}
\end{figure}

To understand how the problem arises it is useful to 
recall how the abelian background field translates from the fundamental
to the symmetric representation. Any given link variable of the SU(3) abelian background field has the form
\begin{equation}
   V^{\rm fun}(x,\mu)= \exp[i\,\diag(\phi^f_1,\phi^f_2,\phi^f_3)],
\end{equation}
and is mapped to a $6\times 6$ diagonal unitary matrix in the symmetric represenation,
\begin{equation}
   V^{\rm sym}(x,\mu)= \exp[i\,\diag(\phi^s_1,\phi^s_2,\phi^s_3,\phi^s_4,\phi^s_5,\phi^s_6)],
\end{equation}
where the angular variables are related as follows:
\begin{equation}
\phi^s_1= 2\phi^f_1,\quad 
\phi^s_2= \phi^f_1+\phi^f_2,\quad 
\phi^s_3= \phi^f_1+\phi^f_3,\quad 
\phi^s_4= 2\phi^f_2,\quad 
\phi^s_5= \phi^f_2+\phi^f_3,\quad 
\phi^s_6= 2\phi^f_3.
\end{equation}
Hence, the fermions in the sextet representation see a background field that is twice 
as large as for fermions in the fundamental representation.
A possible cure could thus be a weaker background field where the boundary gauge fields are halved, 
so that $B\rightarrow B/2$.  Indeed this reduces the cutoff effects to reasonable levels 
for the sextet representation, as seen
in figure~\ref{fig3}.

\section{Conclusions and future directions}

Our perturbative results are a clear warning for anyone embarking on 
a non-perturbative study based on the SF coupling. 
In particular, we find that models with fermions in other than the fundamental
representations can suffer from very large cutoff effects even if these are small
in QCD with otherwise the same parameters\footnote{Similar conclusions regarding the standard SF have been 
reached in~\cite{Karavirta}.}. Note that this problem is not 
cured by a smoothening of the gauge field, as the background field is 
already smooth and thus would not be changed by a smearing procedure.
In particular, our perturbative results are directly relevant for studies such as~\cite{DeGrand:2011vp}.

It seems that models with fermions in other than the fundamental representations call for
a different choice of the background field. Here we have explored a simple minded reduction by a factor
$1/2$. However, while the cutoff effects in the non-fundamental representations are indeed rendered small, 
very large effects are now seen with fermions in the fundamental representation (QCD).
It is not clear to us whether choices exist with universally small cutoff effects. 
Note that any modification of the background field requires a re-assessment of the quality of the signal 
in numerical simulations, as well as a new one-loop calculation for the pure gauge part. 
A more systematic exploration is currently in progress~\cite{progress}.

Finally, our study establishes that the mechanism of automatic O($a$) improvement
works for the SF coupling at one-loop order if regularised with the  $\chi$SF.
However, to mimick the situation of a non-perturbative simulation one should 
determine both $m_{\rm cr}$ and $z_f$ on finite lattices too. This may change the size of 
residual cutoff effects significantly on the small lattices accessible 
to numerical simulations~\cite{progress}.

\subsection*{Acknowledgments}

The authors gratefully acknowledge support by SFI under grant 11/RFP/PHY3218
and by the EU unter grant agreement number PITN-GA-2009-238353 (ITN STRONGnet).

\end{document}